\documentclass{article}

\PassOptionsToPackage{numbers, compress}{natbib}

\usepackage[final]{neurips_mlsb_2021}

\usepackage[utf8]{inputenc} 
\usepackage[T1]{fontenc}    
\usepackage{hyperref}       
\usepackage{url}            
\usepackage{booktabs}       
\usepackage{amsfonts}       
\usepackage{nicefrac}       
\usepackage{microtype}      
\usepackage{xcolor}         
\usepackage{graphicx}
\usepackage{multirow}
\usepackage{amsmath}

\title{Simple End-to-end Deep Learning Model for CDR-H3 Loop Structure Prediction}

\author{%
Natalia Zenkova\\
JetBrains Research\\
\texttt{natalia.zenkova@jetbrains.com} \\
\And
Ekaterina Sedykh \\
University of Tartu \\
\texttt{ekaterina.sedykh@ut.ee} \\
\AND
Tatiana Shugaeva \\
BIOCAD \\
\texttt{shugaeva@biocad.ru} \\
\And
Vladislav Strashko \\
BIOCAD \\
\texttt{strashko@biocad.ru} \\
\And
Timofei Ermak \\
BIOCAD \\
\texttt{ermak@biocad.ru} \\
\And
Aleksei Shpilman\\
HSE University \\
JetBrains Research \\
\texttt{alexey@shpilman.com} \\
}

\begin{document}

\maketitle

\begin{abstract}
Predicting a structure of an antibody from its sequence is important since it allows for a better design process of synthetic antibodies that play a vital role in the health industry. Most of the structure of an antibody is conservative. The most variable and hard-to-predict part is the {\it third complementarity-determining region of the antibody heavy chain} (CDR H3). Lately, deep learning has been employed to solve the task of CDR H3 prediction. However, current state-of-the-art methods are not end-to-end, but rather they output inter-residue distances and orientations to the RosettaAntibody package that uses this additional information alongside statistical and physics-based methods to predict the 3D structure. This does not allow a fast screening process and, therefore, inhibits the development of targeted synthetic antibodies. In this work, we present an end-to-end model to predict CDR H3 loop structure, that performs on par with state-of-the-art methods in terms of accuracy but an order of magnitude faster. We also raise an issue with a commonly used RosettaAntibody benchmark that leads to data leaks, i.e., the presence of identical sequences in the train and test datasets.

\end{abstract}

\section{Introduction}
Antibodies play a key role in adaptive immunity. Their ability to specifically bind target molecules makes antibodies effective fighters against pathogens \cite{tiller2015advances} and one of the major classes of therapeutic molecules in the field of anticancer therapy, autoimmune diseases, and more \cite{ecker2015therapeutic} \cite{reichert2017antibodies}. Experimental determination of the tertiary structure is expensive, therefore, computational approaches are widely used nowadays \cite{tiller2015advances}. Antibodies are composed of two types of protein chains: heavy and light, presented in two copies. The light chain contains variable (VL) and constant domains (CL), the heavy chain contains one variable (VH) and three constant domains (CH1, CH2, CH3). The VH-VL pair forms a variable antigen-binding Fv fragment, which in turn is subdivided into conserved framework regions (FRs) and hypervariable complementarity-determining regions (CDRs). CDRs are the most important regions for antigen binding; they form six variable loops on the surface of the Fv fragment. Three of them are located on the light chain and are named L1, L2, L3, respectively, three others are located on the surface of the heavy chain and are designated as H1, H2, and H3 \cite{marks2017antibody}.

Loops L1-L3 and H1-H2 typically fall into a limited set of conformations, therefore, it is relatively easy to determine their 3D configuration \cite{chothia1987canonical}. In contrast, the third loop of the heavy chain is the most variable and cannot be described using the limited set of structural clusters \cite{north2011new}. Its high variability does not allow the application of modern protein folding techniques based on multiple sequence alignment (MSA) of homologous sequences, such as AlphaFold2 \cite{jumper2021highly} and RoseTTAFold \cite{baek2021accurate}. Thus, CDR H3 structure prediction requires special approaches.

Recently, in addition to statistical and physics-based \cite{sivasubramanian2009toward} prediction methods, deep learning models have been employed for the task of H3 structure prediction. Current state-of-the-art solution DeepH3 uses deep learning to calculate the distribution of inter-residue distances and orientations, which are then passed to the RosettaAntibody package that employs both physics-based and statistical methods. Because of that second step and the causing time constraints, this method can not be feasibly used to perform a wide {\it in silico} screening of possible antibody structures.

In this paper, we present a simple end-to-end deep learning method based on ELMo embedding and forward and backward LSTM passes that directly output coordinates of backbone atoms of amino acids in the H3 loop. This method, that we named \textbf{SimpleDH3}, performs on on par with DeepH3 in terms of root mean square deviation (RMSD) but is at least an order of magnitude faster.


\section{Model}
Our SimpleDH3 model pipeline consists of the following steps:
\begin{enumerate}
    \item Convert each amino acid from the input sequence into a 1024-dimensional embedding representation using a context-dependent model ELMo (Embeddings from Language Models)~\cite{ElMo} pretrained on the UniRef50 dataset \cite{UniRef} from \cite{heinzinger2019modeling};
    \item Independent forward and backward LSTM \cite{LSTM} pass. LSTM cell with 256 neurons in the hidden and the output layer receives an amino acid embedding from ELMo model concatenated with the prediction coordinates for the previous amino acid.
    \item Independent feedforward neural networks with 9-dimensional output after LSTM cell. 9 dimensions correspond to the coordinates of $C$, $C_{\alpha}$ and $N$ atoms in a single amino acid that completely determine its position and orientation. 
    \item Averaging the output from forward and backward passes.
\end{enumerate}
\begin{figure}[h]
    \centering
    \includegraphics[width=\linewidth]{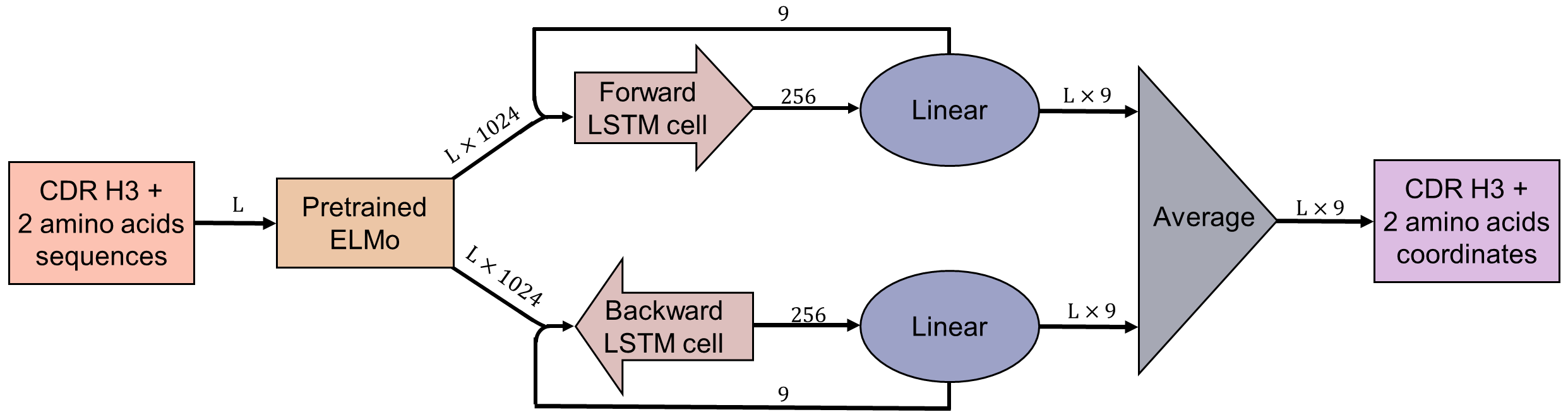}
    \caption{Architecture of our simple model (SimpleDH3). It takes the sequence for CDR H3 and 2 additional amino acids: one on the each end. Each amino acid from a sequence is transformed with pretrained ELMo model from \cite{heinzinger2019modeling}. Afterwards,
    the embeddings are concatenated with the prediction coordinates for the previous amino acid are passed to the backward LSTM cell pass combined with linear layer. Then independently forward LSTM cell pass combined with its own linear layer. As a result, the model outputs backbone coordinates, i.e., for $C$, $C_{\alpha}$, and $N$ atoms, averaged from both passes.}
    \label{fig:model}
\end{figure}

It is important to note that SimpleDH3 model is trained only on CDR H3 loops with two amino acids at the ends. In contrast to DeepH3 \cite{DeepH3} and other methods \cite{DeepAb} which take Fv fragment, our model use only loop sequences.

The model was trained with the following loss function:
\begin{equation*}
    \mathcal{L} = \lambda_1\mathcal{L}_{coords} + \lambda_2\mathcal{L}_{dist} + \lambda_3\mathcal{L}_{torsion}, 
\end{equation*}

where 
\begin{gather*}
    \mathcal{L}_{coords} = MSE_{coords},\ \
    \mathcal{L}_{dist} = MSE_{dist},\\
    \mathcal{L}_{torsion} = MSE_{torsion}, \ \ 
    \lambda_1 = 0.2, \ \
    \lambda_2 = 3.0, \ \
    \lambda_3 = 10.0. 
\end{gather*}

$\mathcal{L}_{coords}$ is a mean-square deviation of $C$, $C_{\alpha}$ and $N$ atoms coordinates; $\mathcal{L}_{dist}$ is a mean-square deviation of distance between two consecutive atoms, such as $(C;C_{\alpha})$,  $(C_{\alpha};N)$ and $(N;C)$; $\mathcal{L}_{torsion}$ is similar to other components of the loss function, but considers torsion angles cosines instead.

\section{Experiments and Results}
\subsection{Data collection and preprocessing}\label{subsec:data_coll}
The training and testing datasets were retrieved from the SAbDab database \cite{sabdab2014}. Only the structures with resolution better than 3.0Å threshold are selected and all sequences are annotated by Chothia numbering scheme \cite{chothia}. Then Fv fragments are extracted using the annotation information. To make the dataset non-redundant, agglomerative clustering is applied \cite{jain1988algorithms} (with the complete linkage rule) with \(1 - sequence\, identity\) as a distance metric (for two chain antibodies, the distance metric is the average mean of \(1 - sequence\, identity\) for heavy and light chains). We break down two-chain (VH-VL) and single-chain (VHH) structures into different bins and clusterize them separately. Next, we determine 100 farthest clusters, i.e., the ones with the maximal sum of the distances to all other clusters, from the two-chain and 50 farthest ones from the single-chain bin with a greedy approach and marked them as the test dataset. The rest of the clusters are the training set. Then, from each cluster we take only the center, which is determined by RMSD among the cluster members, considering the center as the most representative structure of the cluster. Based on the annotation, CDR H3 loop structures are extracted and then only those structures are used for training and testing. Afterwards, we filter out CDR H3 structures containing gaps.

 We take +1 additional amino acids before the loop and +1 after. We then shift the coordinates so that the first atom's coordinates are $(0, 0, 0)$ and the last one is placed on the Y-axis. It is important to note that DeepH3 is trained on a full Fv fragment and our algorithm, in that way, requires less data to train.

\subsection{Train and test split}
The model was trained on 1970 structures from data described in subsection~\ref{subsec:data_coll}.

To compare our results with other methods, we have tested the network on the commonly used RosettaAntibody benchmark \cite{RosettaAntibody}, containing 49 structures with CDR H3 loop lengths ranging from 7 to 17 residues. First, we have deleted structures with these 49 IDs from our dataset. However, we have quickly identified identical CDR H3 loop sequences with different IDs in our training data. The structures are different if compared by the whole Fv-fragment sequences, but if we only measure RMSD on the CDR H3 loop, RMSD on the test dataset will be wrongfully better due to a problem, known as data leaks. Therefore, we eliminated them so there are only unique sequences in all train and test sets. After this step, the size of our train dataset has decreased to 1945 structures. To make a fair comparison of our method to others we do it on the original RosettaAntibody benchmark. However, to make a fair evaluation, we also present the result on the data without duplicate sequences.

\subsection{Training setup}

SimpleDH3 model was trained using NVIDIA Tesla K80 GPU for 500 epochs (~2 hours) until convergence on the antibody dataset, described in Section~\ref{subsec:data_coll} with batch size of 64 and a learning rate of 0.001.

\subsection{Results}

Table~\ref{tbl:results} shows the results of our model in terms of mean RMSD and mean inference time per sequence in seconds.  

As we can see from the results, our simple model performs on par with current state-of-the-art DeepH3 model, but dramatically outperforms it in terms of inference time. Moreover, the time results for DeepH3 \cite{DeepH3} only show the time it takes to run the network part of their method, but, as Rosetta package is needed to get final results, the speed benefit will only increase.

\begin{table}[h]
\centering
\begin{tabular}{|c|c|c|}
\hline
Methods  & Mean RMSD & Mean time per sequence  \\
\hline
DeepH3 \cite{DeepH3}  & 2.2  & 6.65 \\
\hline
SimpleDH3 (ours)  & \textbf{2.04} & \textbf{0.59} \\
\hline
SimpleDH3 (ours, unique) & 2.77 & \textbf{0.59} \\
\hline
\end{tabular}
\vspace{2mm}
\caption{Performance of CDR H3 loops prediction on RosettaAntibody benchmark for both non-unique (original) and unique data.}
\label{tbl:results}
\end{table}

\begin{figure}[h]
    \centering
    \includegraphics[width=1\linewidth]{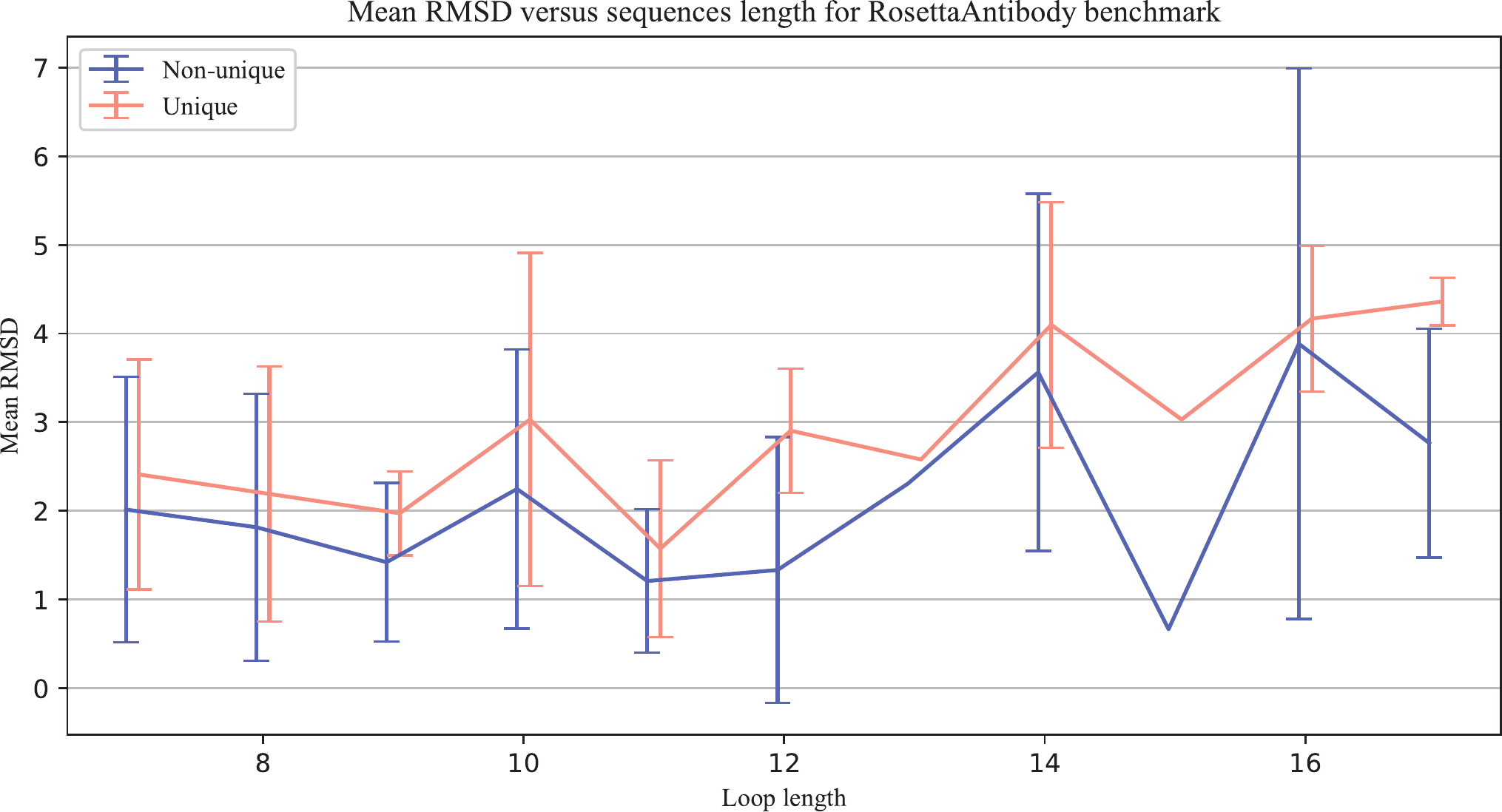}
    \caption{SimpleDH3 RMSD performance across various H3 loop sequence lengths.}
    \label{fig:plot}
\end{figure}

We also show the dependence of RMSD on the sequence length for both non-unique (original) and unique data (Figure~\ref{fig:plot}. RMSD is measured only on target H3 loop sequence, not taking into account additional amino acids used in training and inference.

\section{Conclusion}

In this paper, we present a novel end-to-end deep learning method for the prediction of CDR H3 loop structure. By utilizing pretrained ELMo model and independent forward and backward LSTM passes, we can directly output the coordinates of atoms in the backbone of the amino acid sequence by only training on loop sequences and not whole Fv regions. 

We demonstrate that SimpleDH3 approach performs on par with state-of-the-art methods in terms of RMSD, but is at least an order of magnitude faster. This brings us closer to the possibility of large-scale {\it in silico} screening for a specific antibody for a targeted therapeutic target.

We also point out a flaw in the current benchmark for this task regarding identical sequences with different PDB-IDs and present a method for creating a better benchmark.

Our code is available at \url{https://github.com/jbr-ai-labs/SimpleDH3}.

\newpage
\section*{Acknowledgments}
The publication was supported by the grant for research centers in the field of AI provided by the Analytical Center for the Government of the Russian Federation (ACRF) in accordance with the agreement on the provision of subsidies (identifier of the agreement 000000D730321P5Q0002) and the agreement with HSE University  No. 70-2021-00139. 
\bibliographystyle{abbrvnat}
\bibliography{biblio}

\end{document}